\documentclass[preprint, aps, pre]{revtex4}
\usepackage{graphicx, bm, bbm,amsmath, amssymb, epsfig}

\begin{document}
\newcommand{\nn}{{\noindent}}
\title{Color scales that are effective in both color and grayscale}
\author{Silas Alben}
\affiliation{Department of Mathematics, University of Michigan, \
Ann Arbor, MI 48109} \email{alben@umich.edu}
\date{\today}

\baselineskip=30pt
\begin{abstract}
We consider the problem of finding a color scale which performs
well when converted to a grayscale. We assume that each color is
converted to a shade of gray with the same intensity as the color. 
We also assume that the
color scales have a linear variation of intensity and hue,
and find scales which 
maximize the average chroma (or ``colorfulness'') of the colors. We
find two classes of solutions, which traverse the color wheel in
opposite directions. The two classes of scales
start with hues near cyan and red. The average
chroma of the scales are 65-77\% those of the pure colors. 
\end{abstract}
\maketitle
It is often desirable to present quantitative 
results in both color and grayscale formats. 
In the black-and-white or grayscale format, 
an interval of numerical values maps to a scale of 
grays which increase uniformly in
lightness. Color is usually represented as a three-dimensional
object, and could thus be used to represent a three-dimensional
region of numerical values. More typically, a color scale is
used to represent a one-dimensional interval of numerical
values. Colors are usually easier to distinguish
than grays because of the additional dimensions along which
colors vary. Thus a color scale is generally preferable to
a grayscale.

However, black and white printing is less
costly than color printing, so it is useful to be able to
print color images in black and white and obtain
readable images. Such images are also readable
by those with some degree of color blindness, which 
includes about 10\% of males and a much smaller 
percentage of females.

A color can be defined in different ways. Here we use the
common RGB color model \cite{plataniotis2000color,kang2006computational}. 
A color is defined by a three-component vector, 
$[R,G,B]$, where the components represent the values of red, green, and 
blue in the color. The components $R,G,B$
lie on the interval $[0, 1]$, so a color corresponds to 
a point in the unit cube, $[0, 1]^3$, in RGB space. For example,
pure red is $[1,0,0]$, pure green is $[0,1,0]$, 
and pure blue is $[0,0,1]$. Black is $[0,0,0]$, white is
$[1,1,1]$, and grays are of the form $[g,g,g]$, $0 < g < 1$.

Colors can also be represented in terms of three other quantities:
hue, chroma, and intensity. We define
\begin{align}
M &= \mbox{max}(R,G,B), \\
m &= \mbox{min}(R,G,B). \\
L &= R + G + B - M - m.
\end{align}
\nn Here $L$ is the middle of the components in order of magnitude.
The chroma $C$ is
\begin{align}
C &= M - m,
\end{align}
\nn and represents the colorfulness of a 
given color. For the pure colors already mentioned $C = 1$, and $C = 0$
for any shade of gray. Hue is a cyclical quantity, 
represented by angles $\theta$ ranging from 0 to 360 degrees. Hue is
defined piecewise:
\begin{align}
 \theta &=
  \begin{cases}
   \text{undefined} & \text{if } C = 0 \\
   60\frac{G-B}{C}\; \text{mod}\; 360 & \text{if } M = R \\
   60\frac{B-R}{C} + 120 & \text{if } M = G \\
   60\frac{R-G}{C} + 240 & \text{if } M = B. \label{theta}\\
  \end{cases}
\end{align}
\nn Red corresponds to 0 (and 360) degrees, green to 120 degrees, 
and blue to 240 degrees.
At intermediate angles we have intermediate colors: at 60 degrees
is pure yellow ($[1,1,0]$ in $R,G,B$), at 180 degrees is pure cyan 
($[0,1,1]$), and at 300 degrees is pure magenta ($[1,0,1]$). 

The intensity $I$ (or luma) is a weighted average of $R,G,B$
based on their contribution to the perceived lightness of the color:
\begin{align}
I = 0.299 R + 0.587 G + 0.114 B. \label{I}
\end{align}
\nn The particular weights in (\ref{I}) correspond to 
the National Television System Committee (NTSC) standard,
and have to do with how color is perceived by the eye and
brain. Thus pure blue is darker than pure red, which is darker
than pure green. 

A color scale is represented as a grayscale by converting
each color $[R,G,B]$ to $[I,I,I]$ (from \ref{I}), which yields 
a shade of gray with the same intensity as the color. We
now define certain properties which are desirable for
color scales to perform well both as color scales and as 
grayscales with this conversion.
First, the
intensity should vary linearly from the minimum intensity
$I_{min}$ to the maximum intensity $I_{max}$. Second,
$I_{min}$ should be close to 0 and $I_{max}$ should be close
to 1, to obtain good resolution with the grayscale. Third,
the hues should range over a large portion of $[0, 360]$,
to obtain good hue resolution with the color scale. Fourth,
the hues should increase or decrease monotonically over
the whole color scale to make the colors easy to identify
uniquely by avoiding replication of hues. We further
assume that the hues should increase or decrease 
linearly in the color scale. This assumption is perhaps 
less necessary than linear variation of intensity, but it 
simplifies our calculations by restricting the range of possible 
color scales to a low-dimensional space. Also, the 
chroma of the color scale should be as large as possible on average, to
make colors less gray and thus easier to distinguish.
With these properties
and assumptions, a color scale is represented by curves in both the 
$R,G,B$ and $\theta, C, I$ spaces. As a curve parametrized by $s$, 
$0 \leq s \leq 1$ in ($\theta, C, I$)-space, the color scale satisfies 
\begin{align}
\theta(s) &= \theta_i + s(\theta_f-\theta_i), \label{lintheta}\\
I(s) &= I_{min} + s(I_{max}-I_{min}). \label{linI}
\end{align}
\nn With these choices for $\theta$ and $I$, $C$ is chosen to be as large as
possible on average. By (\ref{theta}), the set of colors with a given hue 
(i.e. $\theta$) is the intersection of a plane with the $R,G,B$ cube. 
The set of $R,G,B$ with a given $I$ is also the intersection of
the plane (\ref{I}) with the $R,G,B$ cube. The intersection of these planes
is a straight line, on which $C$ varies. $C$ is a linear function of $M$
and $m$, and for a given hue, a linear function of $R,G,$ and $B$ also. 
Thus the maximum of $C$ occurs at one of the end points of the line, at
the boundaries of the cube. On the boundaries of the cube, 
either $m$ = 0 or $M$ = 1 (or both). 
To determine which of $m$ = 0 or $M$ = 1 holds, we consider the following
facts. If the hue is fixed, then by (\ref{theta}) so is the ratio $(L-m)/(M-m) = k$.
Imagine we set $m$ = 0 and increase both $L$ and $M$ from zero, keeping the ratio
$(L-m)/(M-m) = L/M$ fixed and equal to $k$. Then the chroma and intensity both
increase monotonically for a fixed hue. The chroma has its maximum when $m = 0,
M = 1,$ and $L = k$. Now imagine we set $M$ = 1 and decrease 
both $L$ and $m$ from 1, keeping the ratio
$(L-m)/(M-m) = (L-m)/(1-m)$ fixed and equal to $k$. The chroma increases
monotonically while the intensity decreases monotonically. The chroma again
has its maximum when $m = 0, M = 1,$ and $L = k$. Thus the set of points
of a given hue and either $m = 0$ and $M = 1$ is a union of two intervals which
meet at a common point with $m = 0, M = 1,$ and $L = k$. The intensity
increases monotonically on this interval and thus equals a given intensity $I$ 
exactly once. If we know both the hue $\theta$ and the intensity $I$, we can determine 
$m, L,$ and $M$ at this point, and also $R, G, B$. We proceed as follows.
We use the hue to find $k$ since $R, G, B$ correspond uniquely to 
$m, L,$ and $M$ for a given hue. Let $I_0$ be the intensity 
when $m = 0, M = 1,$ and $L = k$. If
$I \leq I_0$, we set $m = 0$ and find $L$ and $M$ by solving the two equations
(\ref{I}) and $(L-m)/(M-m) = k$ for the two unknowns $L$ and $M$. Knowing the
hue, we first express (\ref{I}) in terms of $m, L,$ and $M$. If 
$I > I_0$, we set $M = 1$ and find $m$ and $L$ by solving the same two equations.
In this way we obtain the unique $m$, $L$, and $M$ with a given $k$ and $I$ and
with either $m$ = 0 and $M$ = 1. This is the same as the point with a given
hue and intensity for which the chroma is maximized. Having $m$, $L$, and $M$,
and knowing the hue, we can obtain $R$, $G$, and $B$.

Once we have defined $\theta_i$, $\theta_f$, $I_{min}$ and $I_{max}$,
the preceding algorithm defines a color scale. A simple choice is to use the
full ranges of hues and intensities. We also use slightly less than the full ranges
to see how the optimal color scales are affected. 
The range of hues can be traversed
in either the direction of increasing or decreasing $\theta$. Thus we
use  $\theta_f - \theta_i = \pm 360$, as well as the slightly smaller
ranges $\theta_f - \theta_i = \pm 300$ for comparison. For the
intensity range, we use $I_{min} = 0$ or $0.1$ and $I_{max} = 0.9$ or $1$.
These four hue ranges and four intensity ranges yield 16 combinations.
For each combination, we vary $\theta_i$ over $[0, 360]$ and find the
corresponding color scales. We compute the average chroma for 
each color scale:
\begin{align}
\langle C \rangle = \frac{1}{\theta_f-\theta_i}\int_{\theta_i}^{\theta_f} C d\theta.
\end{align}
\nn We define an optimal color scale as that which maximizes
$\langle C \rangle$ over $\theta_i$ for a given 
$\theta_f - \theta_i$, $I_{min}$ and $I_{max}$.

\begin{figure}
  \centerline{\includegraphics[width=7in]
  {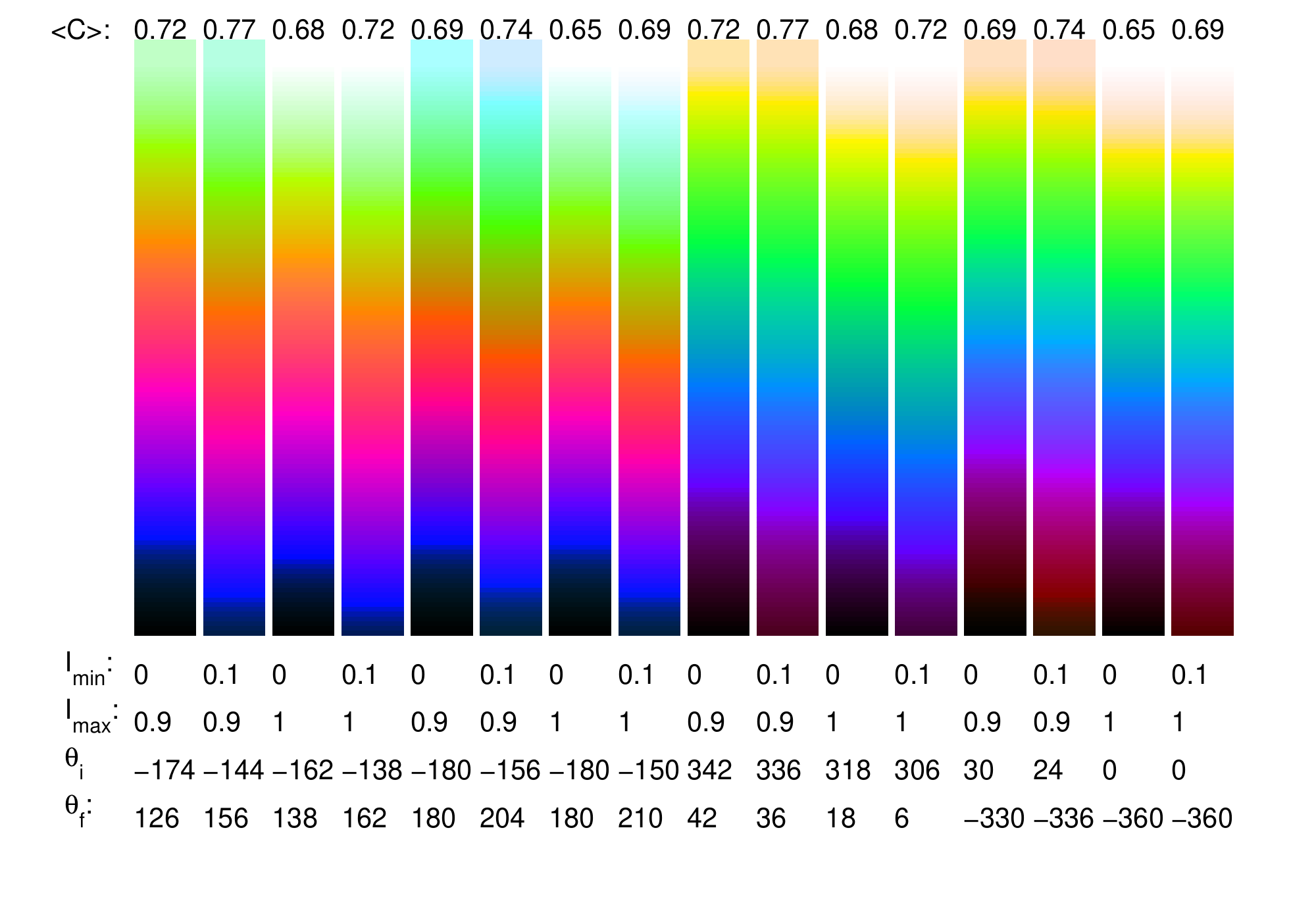}}
  \caption{Color scales which maximize the average chroma $\langle C \rangle$, 
with the assumption of linear variation in hue and intensity, and 16 combinations
of $I_{min}$ (0 or 0.1), $I_{max}$ (0.9 or 1), and $\theta_f - \theta_i$
(-360, -300, 300, or 360). For these 16 parameter sets,
$\langle C \rangle$ is maximized over $\theta_i$.}
\label{fig:MaxChroma}
\end{figure}

Figure \ref{fig:MaxChroma} gives the color scales that maximize
the average chroma for the 16 combinations of parameters. The first
eight start near cyan on the hue scale and move to blue, then
magenta, red, yellow, green and cyan. The second eight move
oppositely around the hue scale, and start at red, followed by 
magenta, blue, cyan, green, yellow, and red. 
The average chroma are the same in corresponding members of
the two sets. 

\begin{figure}
  \centerline{\includegraphics[width=7in]
  {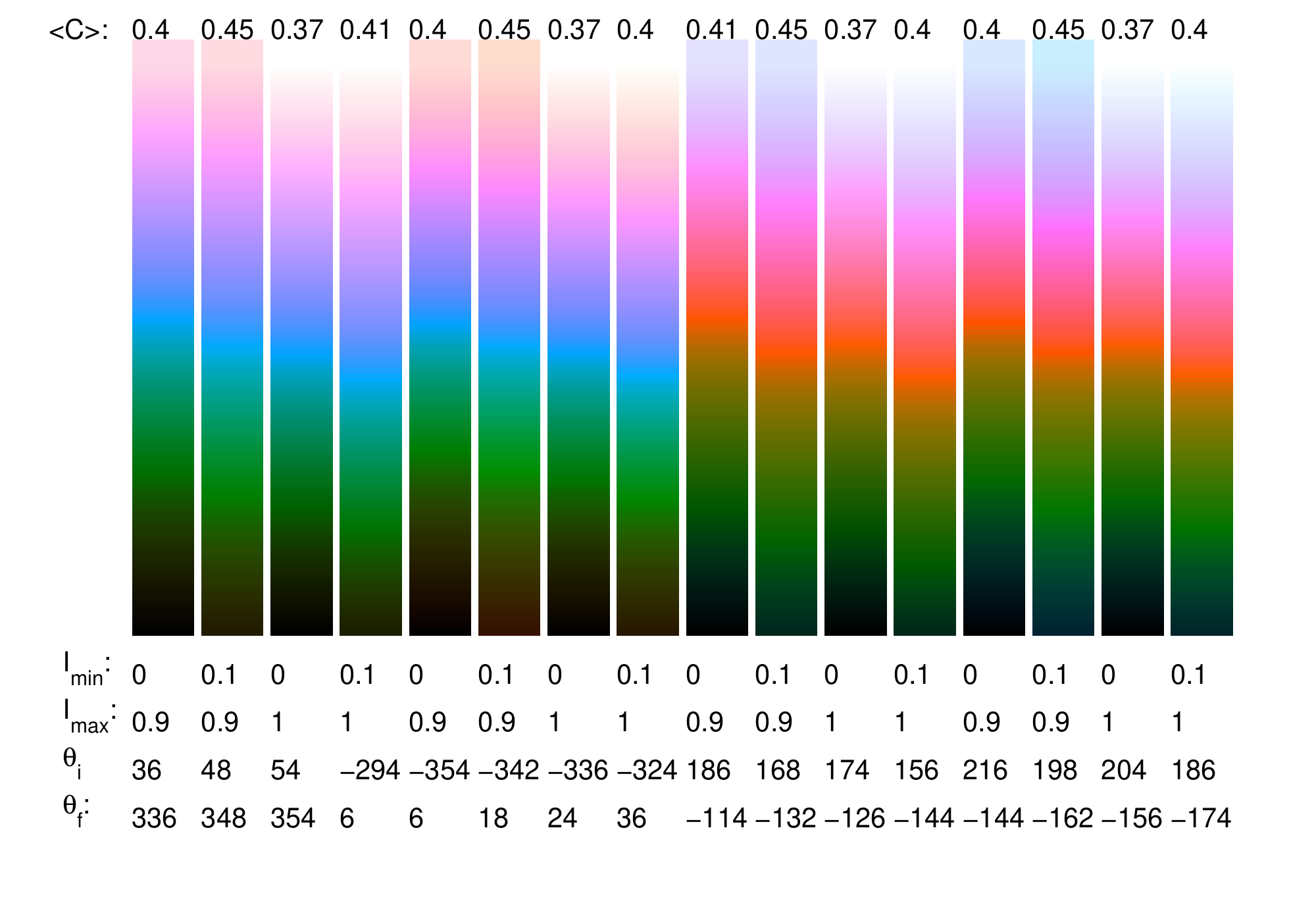}}
  \caption{Color scales which minimize the average chroma over $\theta_i$, for the same
parameters as in figure \ref{fig:MaxChroma}.}
\label{fig:MinChroma}
\end{figure}

It is useful to know how much better the optimal color scales are in 
comparison to non-optimal scales. For a point of reference, 
in figure \ref{fig:MinChroma} we show the chroma-{\it minimizing}
color scales at the same 16 parameters. These color scales are nearly
(but not exactly) opposite in phase with respect to the chroma-maximizing
scales in terms of hue. The first eight start near red, followed by
yellow, green, cyan, blue, magenta, and red. The second eight
start near cyan followed by green, yellow, red, magenta, blue, and cyan.
The average chroma are 55-60\% of those for the optimal scales at
the same parameters. It is apparent that the colors in
figure \ref{fig:MinChroma} are grayer than those in figure
\ref{fig:MaxChroma}.

\begin{figure}
  \centerline{\includegraphics[width=7in]
  {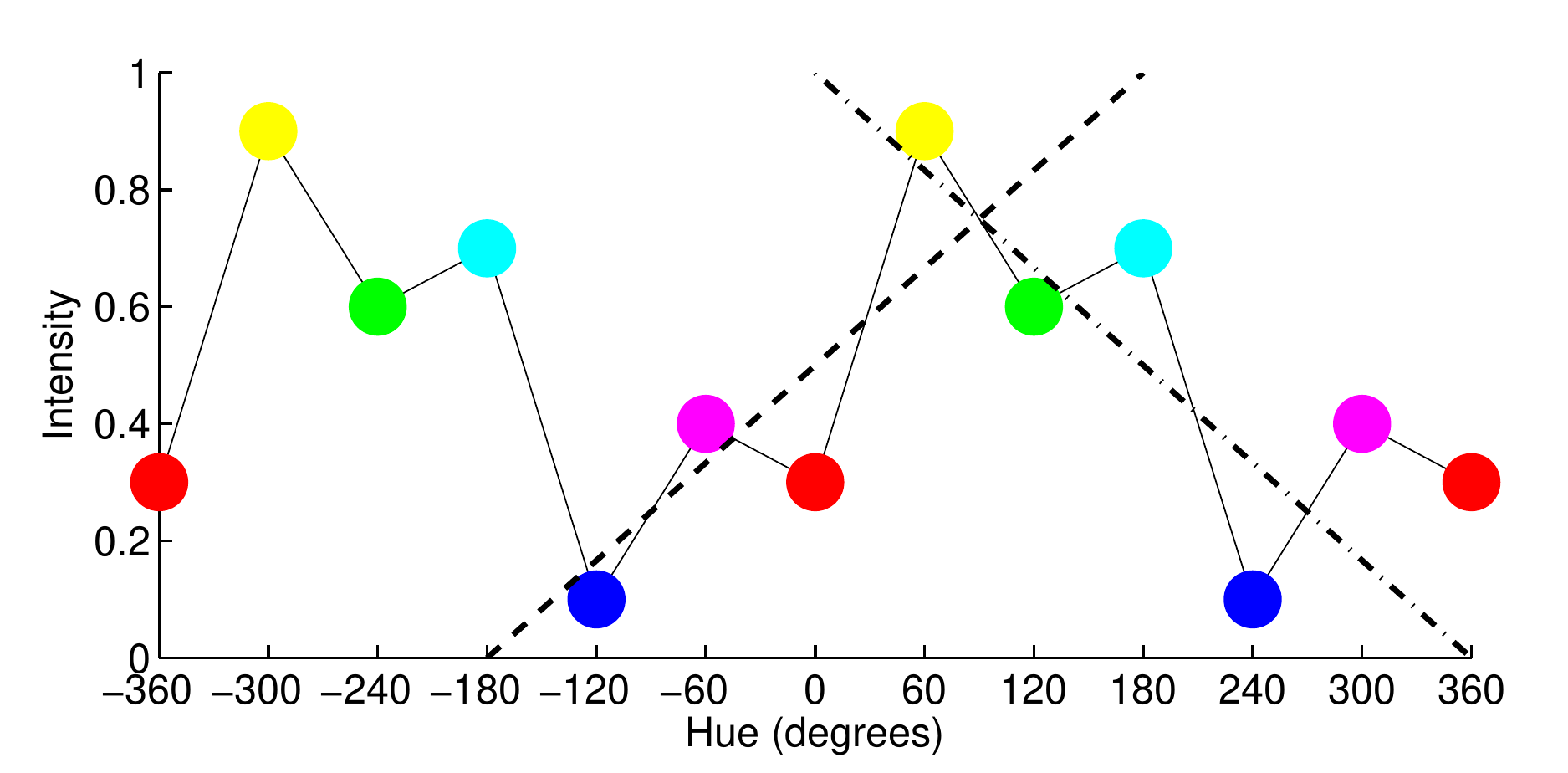}}
  \caption{The chroma-maximizing scales with intensity plotted against hue, for 
$I_{min} = 0$, $I_{max} = 1$, and $\theta_f - \theta_i =$ 360 (dashed line) or
-360 (dashed-dotted line). Also plotted are the intensities of the colors with chroma
of 1 for each hue. These intensities linearly interpolate those of the six pure colors
red, yellow, green, cyan, blue, and magenta.}
\label{fig:Correlation}
\end{figure}

One way of understanding these results is to plot the chroma-maximizing
scales with their intensities versus their hues. We plot 2 of the 16 scales
from figure \ref{fig:MaxChroma} in this way in figure \ref{fig:Correlation}.
At each hue, the intensity of the color with chroma equal to 1 is plotted as
a piecewise linear black solid line which interpolates the six basic colors.
The chroma-maximizing scales are plotted as dashed and
dashed-dotted lines. Shifting these lines horizontally corresponds
to changing $\theta_i$. With the $\theta_i$ shown (-180 and 360 degrees), 
the scales correlate most closely with the chroma-one colors. In both cases,
blue occurs near the low-intensity end of the scale, and yellow near
the high intensity end. Thus the scales incorporate the natural
intensities of the pure colors.


\end{document}